\documentstyle[prl,aps,epsfig,floats]{revtex}
\begin{document}
\draft
\newcommand{\beq}{\begin{equation}}
\newcommand{\eeq}{\end{equation}}
\newcommand{\bea}{\begin{eqnarray}}
\newcommand{\eea}{\end{eqnarray}}
\def\npb#1#2#3{Nucl. Phys. B{#1} (19#2) #3}
\def\plb#1#2#3{Phys. Lett. B{#1} (19#2) #3}
\def\PLBold#1#2#3{Phys. Lett. {#1B} (19#2) #3}
\def\prd#1#2#3{Phys. Rev. D{#1} (19#2) #3}
\def\prl#1#2#3{Phys. Rev. Lett. {#1} (19#2) #3}
\def\prt#1#2#3{Phys. Rep. {#1} C (19#2) #3}
\def\moda#1#2#3{Mod. Phys. Lett.  {#1} (19#2) #3}
\def\lsim{\raise0.3ex\hbox{$\;<$\kern-0.75em\raise-1.1ex\hbox{$\sim\;$}}}
\def\gsim{\raise0.3ex\hbox{$\;>$\kern-0.75em\raise-1.1ex\hbox{$\sim\;$}}}
\def\Frac#1#2{\frac{\displaystyle{#1}}{\displaystyle{#2}}}
\def\al{\alpha}
\def\be{\beta}
\def\ga{\gamma}
\def\de{\delta}
\def\si{\sigma}
\def\C{{\cal{C}}}
\def\O{{\cal{O}}}
\def\wt{\widetilde}
\def\ol{\overline}
\def\l{\left}
\def\r{\right}
\def\no{\nonumber\\}

\twocolumn[\hsize\textwidth\columnwidth\hsize\csname@twocolumnfalse\endcsname

\title{Flavour--dependent SUSY phases and CP asymmetry in $B \to
X_s \gamma$ decays} 
\author{D. Bailin and S. Khalil \\}
\address{Centre for Theoretical Physics, University of Sussex, Brighton BN1
9QJ,~~~U.~K.}
\date{\today}
\maketitle
\begin{abstract}
We study the direct CP asymmetry in $B \to X_s \gamma$ decay in SUSY models with
non--universal $A$--terms. We show that the flavour--dependent phases of the
$A$--terms, unlike the flavour--independent ones, can give rise too large
contribution to the CP asymmetry while respecting the experimental bounds on
the branching ratio of $B \to X_s \gamma$ decay and the electric dipole moments
of the electron and neutron. The CP asymmetry of this decay is predicted to be
much larger than the standard model prediction in a wide region of the
parameter space. In particular, it can be of order $10-15\%$ which can be
accessible at $B$ factories.
\end{abstract}
\pacs{PACS numbers:12.60.Jv, 11.30.Er, 13.20.He 
\hfill SUSX-TH-00-015
}
\vskip.5pc
]
Direct CP asymmetry in the inclusive radiative decay $B \to X_s \gamma$ is measured
by the quantity 
\bea
A_{CP}^{b\to s \gamma} = \Frac{\Gamma(\bar{B} \to X_s \gamma) - \Gamma(B \to
X_{\bar{s}}
\gamma)}{\Gamma(\bar{B} \to X_s \gamma) + \Gamma(B \to X_{\bar{s}}
\gamma)}.
\eea
The Standard Model (SM) prediction for this asymmetry is very small, less than
$1\%$. Thus, the observation of sizeable asymmetry in the decay $B \to X_s \gamma$  
would be a clean signal of new physics.

The most recent result reported by CLEO collaboration for the CP asummetry in  these
decays is \cite{cleo}
\begin{equation}
-9\% < A_{CP}^{b\to s \gamma}  < 42 \%~,
\end{equation}
and it is expected that the measurements of $A_{CP}^{b\to s \gamma}$ will be improved in
the
next few years at the $B$--factories. 

Supersymmetric predictions for $A_{CP}^{b\to s \gamma}$ are strongly dependent on the
flavour structure of the soft breaking terms. It was shown that in the universal case,
as in the minimal supergravity models, the prediction
of the asymmetry is less than $2\%$, since in this case the electric dipole moments
(EDM) of the electron and neutron constrain the SUSY CP--violating phases to be very
small~\cite{Goto,Aoki}. Furthermore, it is also known that in this case, one can not
get any sizeable SUSY contribution to the CP--violating observables, $\varepsilon$ and
$\varepsilon'/\varepsilon$~\cite{Demir,Barr,Abel,Khalil}. The impact of flavour
structure in the non--degenerate
$A$--terms, which is expected in string inspired models, has been
studied~\cite{Abel,Khalil,Vives}. In these works, it was emphasized that the
flavour--dependent SUSY CP--violating phases are not essentially constrained by the
EDMs and can lead to large contributions to the observed CP--violating phenomena in
the kaon system and, in particular, to the direct CP violation measured by
$\varepsilon'/\varepsilon$. 

In this letter, we explore the effect of these large flavour--dependent phases on
inducing a direct CP violation in $B \to X_s \gamma$ decay. We will show that the
values of the asymmetry $A_{CP}^{b\to s \gamma}$ in this class of models are much
larger than the SM prediction in a wide region of the parameter space allowed by
experiments, namely the EDM experimental limits and the bounds on the branching ratio
of $B \to X_s \gamma$ . The enhancement of $A_{CP}^{b\to s \gamma}$ is due to the
important contributions from gluino--mediated diagrams, in this
scenario, in addition to the usual chargino and charged Higgs contributions. 

The effective Hamiltonian for $b \to s \gamma$ decay is given by
\beq
H_{eff}=-\frac{4G_F}{\sqrt{2}}V_{ts}^{*}V_{tb}\sum_{i=1}^{8}
C_i(\mu_b) Q_i(\mu_b)
\label{Heff}
\eeq
where $ Q_i(\mu_b)$ and $ C_i(\mu_b)$ represent an operator for the $\Delta B=1$ transition 
and its Wilson coefficient, respectively, evaluated at the renormalization scale $\mu_b \simeq 
{\cal O}(m_b)$. The relevant operators for $b \to s \gamma$ decay are given by
\bea
Q_2 &=& \bar{s}_L \gamma^{\mu} c_L~ \bar{c}_L \gamma_{\mu} b_L, \no
Q_7 &=& \Frac{e}{16 \pi^2} m_b \bar{s}_L \sigma^{\mu \nu} b_R F_{\mu \nu}, \no
Q_8 &=& \Frac{g_s}{16 \pi^2} m_b \bar{s}_L \sigma^{\mu \nu} T^a b_R G^a_{\mu \nu}.
\eea

The expression for the asymmetry  $A_{CP}^{b\to s \gamma}$, corrected to 
next--to--leading order is given by \cite{Neubert}
\bea
A_{CP}^{b\to s \gamma} &=& \Frac{4\al_s(m_b)}{9 \vert C_7 \vert^2} \biggl\{
\biggl[\frac{10}{9} - 2 z~ (v(z)+b(z,\delta))\biggr] Im[C_2 C_7^*]\no 
&+& Im[C_7 C^*_8] + \frac{2}{3} z~b(z,\delta) Im[C_2 C_8^8] \biggr \}, 
\label{asymmetry}
\eea
where $z=m_c^2/m_b^2$. The functions $v(z)$ and $b(z,\delta)$ can be found in
Ref.\cite{Neubert}. The parameter $\delta$ is related to the experimental cut on the photon
energy, $E_{\gamma} > (1-\delta) m_b/2$, which is assumed to be 0.9. We neglect the
very small effect of the CP--violating phase in the CKM matrix.

The SUSY contributions to the Wilson coefficients
$C_{7,8}$ are obtained by calculating the $b\to s \gamma$ and $b \to s g $
amplitudes at the electroweak (EW) scale respectively. 
The leading--order contributions to these amplitudes are given by the 1--loop
magnetic-dipole and chromomagnetic dipole penguin diagrams respectively,
mediated by charged Higgs boson, chargino, gluino, and neutralino exchanges. 
As pointed out in Ref.\cite{Emidio}, SUSY models with non--universal $A$--terms may induce
non--negligible contributions
to the dipole operators $\tilde{Q}_{7,8}$ which have opposite chirality to $Q_{7,8}$. 
In the MSSM these contributions are suppressed by terms of order ${\cal O}(m_s/m_b)$
due to the universality of the $A$--terms. However, in our case we should take them into
account. Denoting by $\tilde{C}_{7,8}$ the Wilson coefficients multiplying the new
operators $\tilde{Q}_{7,8}$ the expression for the asymmetry in Eq.(\ref{asymmetry})
will be modified by making the replacement 
\beq
C_i C_j^* \to C_i C_j^* + \tilde{C}_i \tilde{C}_j^*.
\label{chirality}
\eeq
The expressions for $\tilde{C}_{7,8}$ are given in
Ref.\cite{Emidio} and $\tilde{C}_2=0$ (there is no operator similar to $Q_2$
containing right--handed quark fields).

In this case the general parametrization of the branching ratio in terms   
of the new physics contributions takes the form~\cite{Emidio,Kagan}
\bea
&&\hspace{-0.5cm}BR(B\to X_s\gamma) = (3.29\pm 0.33)\times 10^{-4}~
\biggl(1 + 0.622 Re[\xi_7] \no
&+& 0.090\bigl(~\vert \xi_7 \vert^2 + \vert \tilde{\xi}_7 \vert^2~\bigr) 
+ 0.066~Re[\xi_8] + 0.019 \bigl(~Re[\xi_7 \xi_8^*] \no
&+& Re[\tilde{\xi}_7 \tilde{\xi}_8^*]~\bigr) + 0.002 \bigl(~
\vert \xi_8 \vert^2~+~\vert \tilde{\xi}_8 \vert^2~\bigr)~\biggr)~,
\label{bsgPAR}
\eea
where $\xi_{7,8}=\bigl(C_{7,8}-C^{SM}_{7,8}\bigr)/C_{7,8}^{SM}$ and 
$\tilde{\xi}_{7,8}=\tilde{C}_{7,8}/C_{7,8}^{SM}$. $C_{7,8}$ include the total
contribution while $C_{7,8}^{SM}$ contains
only the SM ones.

Note that including these modifications (\ref{chirality})
may enhance the branching ratio of $B \to X_s \gamma$ and reduce the CP asymmetry, 
since $\vert C_7 \vert^2$ is replaced by $\vert C_7 \vert^2 + \vert \tilde{C}_7
\vert^2$ in the denominator of Eq.(\ref{asymmetry}). If so, neglecting this
contribution could lead to an incorrect conclusion. 

As already mentioned, usually the SUSY trilinear terms are defined at the high energy 
scale to be proportional to their associated Yukawa matrix, {\it i.e.} 
$Y^A_{ij} = A_0 Y_{ij}$ and $A_0$ is the same for all Yukawa couplings. 
Despite the simplicity of this assumption, it is a very peculiar case
and there exist classes of model inspired by string theory that lead to non--universal
$A$--terms. Moreover, this non--universality is crucial in order to generate the
experimentally observed CP violation $\varepsilon$ and $\varepsilon'/\varepsilon$ even
with the phase $\delta_{CKM}$ vanishing~\cite{Abel,Khalil,Vives}. 

Motivated by these models~\cite{Khalil} we parameterize the
trilinear matrices $Y^A_d$ as $(Y_d^A)_{ij} = A^d_{ij} Y^d_{ij}$, where
$A^d_{ij}$ is
given by
\beq
A^d_{ij} = \left ( \begin{array}{ccc}  
a & a & b\\
a & a & b \\
b & b & c
\end{array} \right).
\label{Atex}
\eeq
The entries $a,b$ and $c$ are complex and of order the gravitino mass, ${\cal
O}(m_{3/2})$. In general $A^d$ and $A^u$ have different structure. Here we
assume for simplicity that $A^d = A^u $. The results, as we will
show, are sensitive only to the non--universality between the entries $A_{ij}$.
Also we assume that the scalar masses and the gaugino masses are universal
and of order $m_{3/2}$ too. So that after rotating the phase of the gaugino masses we
have three phases in the $A$--sector: $\phi_a$, $\phi_b$ and $\phi_c$, which are
the relative phases between the gaugino phase and the original phases of the
$A_{ij}$ entries. It is important to note that the EDM limits on the electron and the
neutron essentially constrain the phase of the entry $A_{11}$ ($\phi_a$) and
the other two phases are much less constrained. These two phases gives large
SUSY contribution to the CP observables. 

We found that the flavour--independent phase $\phi_c$ gives a large contribution to
$\tilde{C}_7$ and $\tilde{C}_8$. This can simply understood by using the mass insertion,
the gluino contributions of $\tilde{C}_7$ and $\tilde{C}_8$ are proportional to
$(\delta^d_{LR})_{23} \simeq (S_{D_L} Y_d^{A*} S^{\dag}_{D_R})_{23}/m^2_{\tilde{q}}$
and 
$(\delta^d_{RL})_{23} \simeq (S_{D_R} Y_d^{A} S^{\dag}_{D_L})_{23}/m^2_{\tilde{q}}$,
where
$ S_{D_R} Y_d S_{D_L}^{\dag} =Y_d^{diag}$. 
It is obvious that the entries $A_{33}$ could give significant contribution to
$(\delta^d_{LR})_{23}$ and $(\delta^d_{RL})_{23}$. As explained above, enhancing the
Wilson coefficients $\tilde{C}_{7,8}$ increases the value of the branching ratio and 
decreases the value of the CP asymmetry.

Indeed, we found that for small values $\phi_c$ the branching ratio is very much
enhanced and becomes above the experimental upper bound, and the CP asymmetry is
reduced a lot. Thus $\phi_c$ is constrained to be very small ($ \lsim {\cal
O}(10^{-1}$)), and
since we are interested in studying scenarios in which the CP asymmetry is sizeable, we
assume that $\phi_c$ is zero\footnote{In any case, we expect that the EDM of the mecury
atom provides a severe constraint on this phase~\cite{Bailin}}. In other words, the
flavour--diagonal phases have to be aligned with the gaugino phase. In fact, this would
be true in some string inspired models where the dilaton vacuum expectation value
(VEV) gives the main contribution to the gaugino masses and the diagonal entries of the
$A$--terms, while the VEVs of the moduli fields give the contribution to the
off--diagonal elements of $A$--terms~\cite{Khalil}.

The effect of the other flavour--independent phase, $\phi_a$ on the $A_{CP}^{b\to s
\gamma}$ and $BR(b \to s \gamma)$ is
found to be very small. So any constraint on this phase from the EDM is not a problem in
this class of models. 

It is worth mentioning that the textures of the $A$--terms that derived from type I
string inspired models are not favoured here since they can not enhance the CP
asymmetry over that in the minimal supergravity models. These textures have the
following form~\cite{Vives}
\beq
A_{ij} = \left ( \begin{array}{ccc}
a & b & c\\
a & b & c \\
a & b & c
\end{array} \right),
\label{Atex2}
\eeq
which means that we can write the parameters $Y^A_{ij}$ in matrix notation as 
$$Y^A = Y~ .~ \left ( \begin{array}{ccc}
a & 0 & 0\\
0 & b & 0 \\
0 & 0 & c   
\end{array} \right).$$ 
Thus, working in the SCKM basis where the unitary matrices $S_{L,R}$ are obtained by 
superfield rotation so that the quark mass matrices are diagonal, the trilinear terms become
$ Y_{diag} . A'$ where $A'$ is given by $ A' = S_L A S_L^{\dag}$. 
This result is very similar to the case of the universal $A$--terms. Since the Yukawa 
couplings preserve the diagonality once this basis is chosen at the GUT scale, the effect of 
the off--diagonal element in this model is diluted.
Moreover, as seen from Eq.(\ref{Atex2}), the phases of the elements $A_{22}$ and $A_{33}$
(which are constrained by the EDM and the $BR(b\to s \gamma)$) are the same phases of the 
off--diagonal elements $A_{32}$ and $A_{23}$ respectively. Therefore, this kind of texture  
does not give a significant contribution to the CP asymmetry in the $B \to X_s \gamma$,
as pointed out in Ref.~\cite{Masiero}, although it gives important contribution to the
CP--violating effects in the kaon system, as shown in Ref.~\cite{Vives}.

In the class of models we consider here, Eq.(\ref{Atex}), the relevant and important phase 
for the CP asymmetry is the phase of the off--diagonal elements $\phi_b$. 
In Fig.1 we show the dependence of $A_{CP}^{b\to s \gamma}$ on
$\phi_b$ for $m_{3/2} =150$ GeV and $\tan\beta=3$ and $10$. The non--universality
among $\vert a \vert$, $\vert b \vert$ and $\vert c \vert$ is crucial to get these
results. Large $\vert c \vert$ is required for obtaining the $BR(b\to s \gamma)$ values
within the experimental limits, especially at low $\tan \beta$. This is because 
large $\vert c \vert \equiv A_t$ leads to large off--diagonal elements in the stop
mass matrix. Hence the splitting of the two stop mass eigenstates becomes large
and this enhances the chargino contribution which decreases the total branching
ratio~\cite{Shafi}.

Therefore, for $\tan \beta =10$ we fixed $\vert c \vert \simeq
\sqrt{3} m_{3/2}$ and for $\tan \beta =3$ we used $\vert c \vert \simeq 3~m_{3/2}$
otherwise the values of the branching ratio of $B \to X_s \gamma$ decay are larger
than $4\times 10^{-4}$. Also we observe that $\vert b \vert < \vert c \vert $ is
favoured by large CP asymmetry $A_{CP}^{b\to s \gamma}$.     

\begin{figure}
\epsfig{file=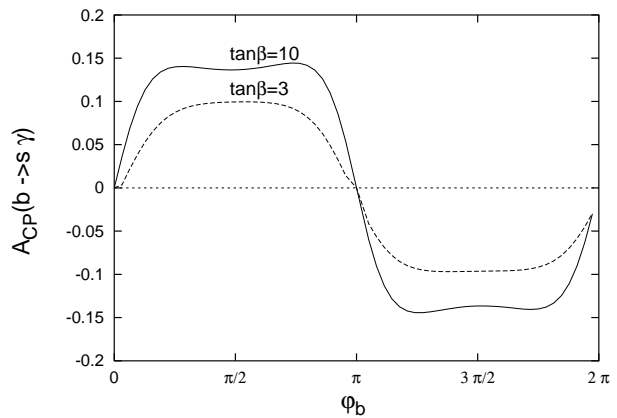,width=8cm}\\

\caption{CP asymmetry $A_{CP}^{b\to s \gamma}$ as a
function of the flavour--dependent phase $\phi_b$, for $m_{3/2} \simeq 150$ GeV and 
$\tan \beta =3$ and $10$.}
\label{fig1}
\end{figure}
We see from Fig.1 that the CP asymmetry $A_{CP}^{b\to s \gamma}$ can be as
large as $\pm 15\%$, which 
can be accessible at the B--factories. Also this result does not require a light chargino 
as in the case considered in Ref.~\cite{Ko}.

It is important to emphasize that the gluino contribution in this model gives the
dominant contribution to the CP asymmetry $A_{CP}^{b\to s \gamma}$. We found
that although the real parts of the gluino contributions to both of $C_{7,8}$ and
$\tilde{C}_{7,8}$ are smaller than the real parts of the other
contributions (but not negligible as in the case of universal $A$--terms), their
imaginary parts are dominant and give with the imaginary parts
of the chargino contribution the main contributions to $A_{CP}^{b\to s \gamma}$. It
is clear that these contributions vanish for $\phi_b$ equal to a multiple of $\pi$
and $A_{CP}^{b\to s \gamma}$ in this case is identically zero as Fig.1 shows.

In Fig.2 we show the correlation between the CP asymmetry $A_{CP}^{b\to s \gamma}$ and
the branching ratio $BR(B\to X_s \gamma)$. We scan over the following region of the 
parameter space: 
$$\tan \beta ~\simeq~ 2-40,~~~ m_0,~m_{1/2},~ \vert a \vert,~\vert b \vert ,~
\vert c \vert~ \simeq~ 100-500~ GeV,$$
where $m_0$ is the common scalar mass and $m_{1/2}$ is the common gaugino
masses. Assuming non--universal scalar and gaugino masses could also help in
enhancing the CP asymmetry, however, here we assume these masses are universal to 
maximizing the effect of the non--universality of the trilinear couplings. 

The phases $\phi_a$ and $\phi_c$ are assumed to be zero while $\phi_b$
is fixed to be $\pi/2$ which gives, as shown in Fig.1, nearly maximum CP asymmetry.
The low energy SUSY spectrum is calculated by running these
soft SUSY breaking terms from the GUT scale $(\simeq 2\times 10^{16}$ GeV)
to the EW scale ($\simeq M_Z$) and imposing all the usual constraints, like the
electroweak symmetry  breaking, absence of colour and electric charge breaking minima,
experimental bound on the SUSY particles, etc.  

\begin{figure}
\epsfig{file=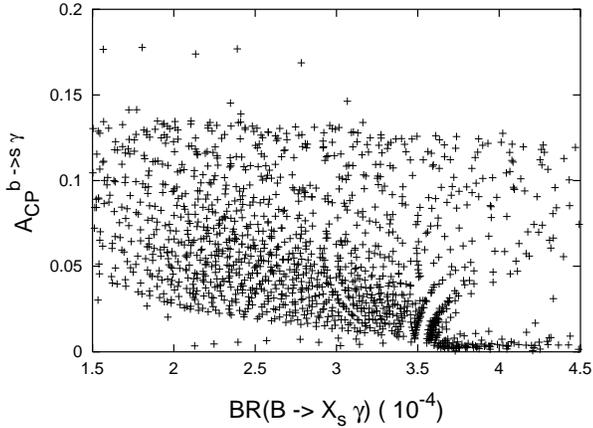,width=8cm}\\

\caption{Correlations of CP asymmetry $A_{CP}^{b\to s \gamma}$ with the branching
ratio $BR(B\to X_s \gamma)$. }
\label{fig2}
\end{figure}  
From this figure we see that there is a wide region of the parameter space where the
decay rate asymmetry of $B\to X_s \gamma$ can be large. We also note that large
values of CP asymmetry $A_{CP}^{b\to s \gamma}$ prefer small values for the branching
ratio $BR(B\to X_s \gamma)$. This correlation is also found in Ref.~\cite{Aoki}.
We fixed the sign of $\mu$ to be positive (in our convention, where the sign of
$\mu$ in chargino mass matrix is positive and in neutralino mass matrix is negative) 
since $\mu <0$ leads to large values of $BR(B\to X_s \gamma)$ and almost the whole
range of
parameter space is excluded. 
\\

In summary, we have considered the possible supersymmetric contribution
to CP asymmetry in the inclusive $B\to X_s \gamma$ decay in model with
non--universal $A$--terms. Contrary to the universal scenario, we
find that the CP asymmetry in this class of models is predicted to be large 
in sizeable regions of the parameter space allowed by the experimental bounds,
and may be possibly to be detected at B factories  We have shown that the
flavour--dependent phases
are crucial for this enhancing with respecting the severe bounds on the electric
dipole moment of the neutron and electron. 
\\

\acknowledgements
S.K. would like to thank S. Abel, F. Borzomati and M. Gomez for useful discussions.
This work was supported by the PPARC.


\end{document}